\documentclass{article}
\usepackage{multirow}
\usepackage{spconf,amsmath,epsfig}
\usepackage[tight,footnotesize]{subfigure}
\usepackage[ruled,vlined]{algorithm2e}

\renewenvironment{thebibliography}[1]{%
   \begin{oldthebibliography}{#1}%
     \setlength{\itemsep}{-.35ex}%
}%
{%
   \end{oldthebibliography}%
}

\DeclareGraphicsExtensions{.png, .PNG, .jpg, .jpeg, .pdf, .PDF}

\title{SemI2I: Semantically Consistent Image-to-Image Translation for Domain Adaptation of Remote Sensing Data}

\name{Onur Tasar$^1$, S L Happy$^2$, Yuliya Tarabalka$^3$, Pierre Alliez$^1$}
\address{$^1$Universit{\'e} C{\^o}te d'Azur, Inria, TITANE team; $^2$Inria, STARS team; $^3$LuxCarta Technology \\
Email: onur.tasar@inria.fr
}

\begin{document}
%
\maketitle
\begin{abstract}
\vspace{-1mm}
Although convolutional neural networks have been proven to be an effective tool to generate high quality maps from remote sensing images, their performance significantly deteriorates when there exists a large domain shift between training and test data. To address this issue, we propose a new data augmentation approach that transfers the style of test data to training data using generative adversarial networks. Our semantic segmentation framework consists in first training a U-net from the real training data and then fine-tuning it on the test stylized fake training data generated by the proposed approach. Our experimental results prove that our framework outperforms the existing domain adaptation methods.
\end{abstract}
\begin{keywords}
Domain adaptation, data augmentation, semantic segmentation, dense labeling, image-to-image translation, generative adversarial networks, GANs.
\end{keywords}
\vspace{-3mm}
\section{Introduction}
\vspace{-3mm}
Over the years, U-net~\cite{ronneberger2015u} and its variants~\cite{khalel2019multi} have become one of the most commonly used network architecture for semantic segmentation problem because of their continuous success in various benchmarks. However, the difference between the data distributions of training and test images caused by atmospheric effects, position of the sun, intra-class variability, etc. makes the U-net likely to fail to generate good maps. To overcome this issue, the traditional data augmentation methods such as gamma correction, random contrast change~\cite{tasar2019incremental, tasar2019continual}, histogram matching~\cite{Gonzalez}, color constancy algorithms~\cite{buchsbaum1980spatial} are widely used. However, adopting such augmentation techniques is usually insufficient to generalize the model to unseen data, especially when the domain shift between training and test data is highly large.

Better data augmentation strategy would be to generate semantically consistent and test stylized fake training data. To accomplish this task, especially in the field of computer vision, various image-to-image translation (I2I) approaches have already been proposed~\cite{zhu2017unpaired, liu2017unsupervised, huang2018multimodal, lee2018diverse}. The biggest challenge for the style transfer problem is to generate test stylized fake training image that semantically matches the real training image, so that one can use the fake training image and the ground-truth for the real training image to train or to fine-tune a model. The recent I2I approaches~\cite{zhu2017unpaired, liu2017unsupervised, huang2018multimodal, lee2018diverse} fail to keep the real and the fake training images semantically the same. For instance, some of them replace trees by buildings, buildings by roads, etc., whereas some of them create totally artificial objects in the fake training image, which are not available in the real training data. Thus, the fake training images generated by these approaches and the ground-truth for the real training data do not correspond. Hence, training a model on such image and ground-truth pair causes the model to yield a poor performance. Another approach is to adapt the model to test data~\cite{tsai2018learning}, which is a more difficult task, rather than modifying the data.

To overcome the limitations described above, Tasar~\textit{et~al.} have recently proposed ColorMapGAN~\cite{tasar2019colormapgan}, which learns to change the color distribution of training data without doing any structural change. Although the method works very well, it has some limitations. Firstly, it does not support style transfer when the images in both domains have different number of channels. Secondly, it learns to map the colors of the source domain linearly, which may not be strong enough in some cases. Finally, since it transforms each color separately, its output is usually slightly noisy. In this work, we propose semantically consistent image-to-image translation (SemI2I) method, which overcomes all the limitations of ColorMapGAN. By following the same segmentation strategy explained in~\cite{tasar2019colormapgan}, we use the fake training images and the original ground-truth to fine-tune the U-net trained on real data.


\begin{figure*}
\centering
    \includegraphics[width=0.9\linewidth]{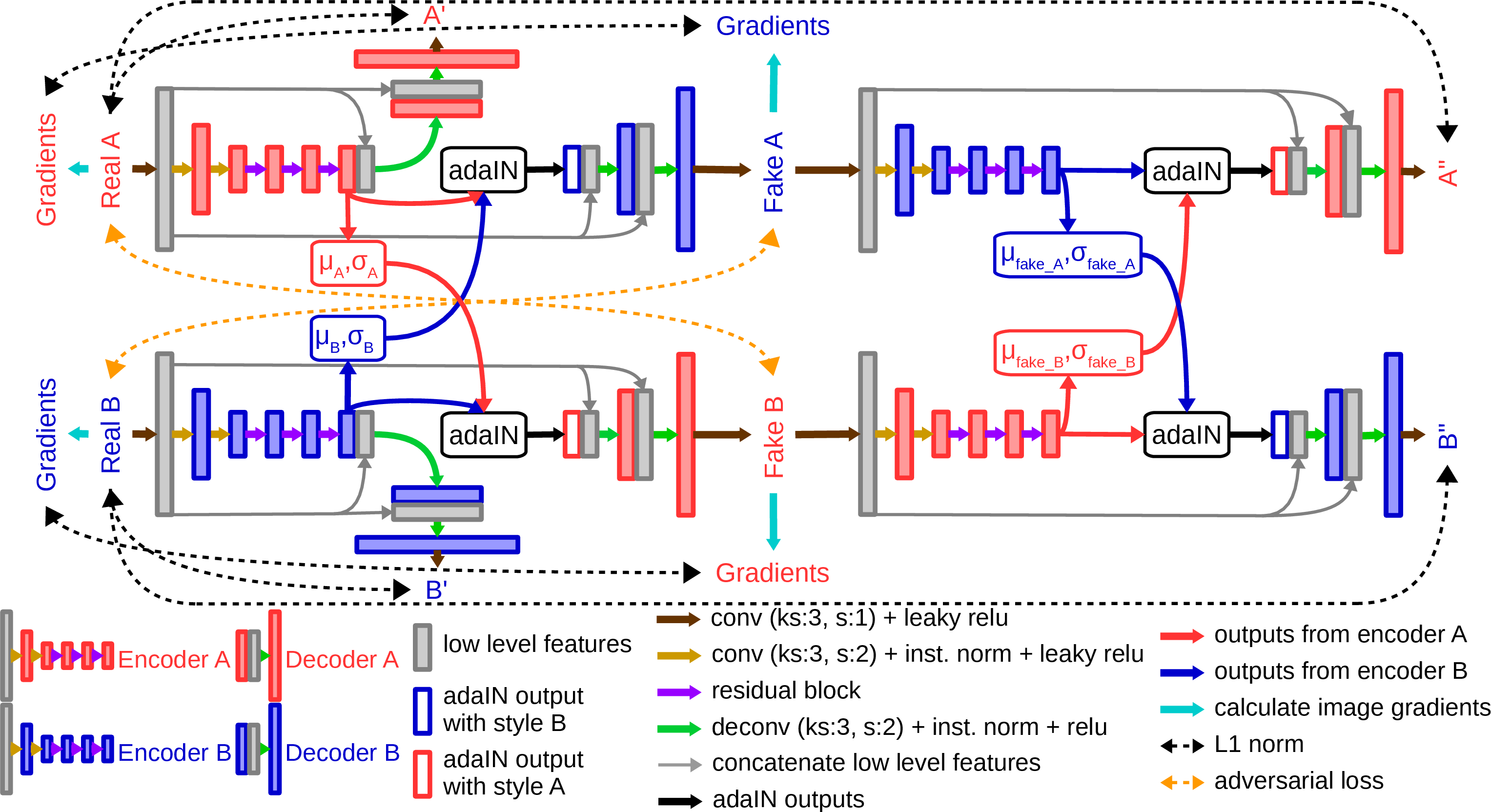}
\caption{SemI2I. $ks$ and $s$ stand for kernel size and strides. $\mu$ and $\sigma$ are the mean and standard deviation values for each feature channel of the embeddings. Since SemI2I performs I2I, one can consider A as training and B as test data, or vise versa.}
\label{fig:method}
\end{figure*}

\vspace{-4mm}
\section{Methodology}
\vspace{-3mm}
Let us assume that both domains are denoted by A and B, respectively. The essential goal of SemI2I is to generate fake A with the style of B, and fake B with the style of A. The design of SemI2I is illustrated in Fig.~\ref{fig:method}. The main components and the stages of SemI2I are as follows:

\textbf{\textit{Style transfer.}} The crucial components for the style transfer are adaptive instance normalization (adaIN~\cite{huang2017arbitrary}) and adversarial losses. To generate B stylized fake A and A stylized fake B, we first encode both A and B. In the encoded embeddings, we compute mean $\mu$ and standard deviation $\sigma$ for each feature channel. The embedding itself has the content information, whereas $\mu$ and $\sigma$ carry the style information. We combine the content of A and the style of B, and the content of B and the style of A by using adaIN. After this operation, since we switch the styles of A and B, we obtain fake A by the decoder B and fake B by the decoder A. Moreover, to make B and fake A, and A and fake B visually as similar as possible, we minimize the adversarial losses between them. The combination of an encoder and a decoder forms a generator. For the discriminator, we use the same architecture as in CycleGAN~\cite{zhu2017unpaired}. As can be verified from Fig.~\ref{fig:method}, there are 6 generators and 2 discriminators in SemI2I.

\textbf{\textit{Semantic consistency.}} By following the same strategy described above, we switch the styles of fake A and fake B to obtain A$''$ and B$''$ (cross reconstruction). In an ideal transformation A and A$''$, and B and B$''$ have to be exactly the same. In addition, when we encode A with encoder A and decode the embedding with decoder A, we must reconstruct itself. We call this operation self reconstruction. The self reconstruction also applies to B. Besides, we compute the image gradients from A and fake A by sobel filter. We force the difference between the gradients calculated from A and fake A to be as small as possible to help them to be semantically consistent. Again, the same rule applies to B. Finally, we have visually verified that the filters in the first convolution layer of each encoder learns the low level features such as edges. We re-size and concatenate these low level features from each encoder with the input to each deconvolution layer in the corresponding decoder (see gray arrows in Fig.~\ref{fig:method}). Such concatenation allows the decoder to have a footprint of the objects in the real data; therefore, it guides the decoder to place the right objects in the correct locations. 

\textbf{\textit{Training.}} We now formulate the losses to train SemI2I. We compute the cross reconstruction loss $\mathcal{L}_{cross}$, which is the sum of L1 norm between A and A$''$ and L1 norm between B and B$''$. Similarly, the self reconstruction loss $\mathcal{L}_{self}$ is computed by summing L1 norms between A and A$'$, and B and B$'$. By adding L1 norms between the image gradients calculated from A and fake A, and the gradients from B and fake B, we compute the gradient loss $\mathcal{L}_{grad}$. Finally, we use the loss functions in LSGAN~\cite{mao2017least} to compute the adversarial losses. The adversarial loss for the generators $\mathcal{L}_{adv\_G}$ is the sum of the LSGAN generator losses between A and fake B, and B and fake A. Similarly, the adversarial loss for the discriminators $\mathcal{L}_{adv\_D}$ is computed by adding the LSGAN discriminator losses between A and fake B, and B and fake A. The overall generator loss $\mathcal{L}_{G}$ that is minimized in the training stage is computed as:
\begin{equation}\label{eq:g_loss}
\mathcal{L}_{G} = \lambda_{1} \mathcal{L}_{cross} + \lambda_{2} \mathcal{L}_{self} + \lambda_{3} \mathcal{L}_{grad} + \lambda_{4} \mathcal{L}_{adv\_G}, 
\end{equation}
where $\lambda_{1}, \lambda_{2}, \lambda_{3}$, and $\lambda_{4}$ are used to adjust the relative importance of each loss. The discriminator loss is defined as :
\begin{equation}\label{eq:d_loss}
\mathcal{L}_{D} = \lambda_{4} \mathcal{L}_{adv\_D}.
\end{equation}
To train SemI2I, we simultaneously minimize $\mathcal{L}_{G}$ and $\mathcal{L}_{D}$.

\textbf{\textit{Test.}} In the test stage, to generate B stylized fake A, we need the encoder A and the decoder B (see the top left generator in Fig.~\ref{fig:method} that generates fake A). However, as can be seen in Fig.~\ref{fig:method}, before feeding the embedding encoded by the encoder A from real A to the decoder B, it needs to be normalized by adaIN using $\mu_B$ and $\sigma_B$ calculated from the embedding of B. When generating fake A, one may think of randomly sampling a patch from B, encoding it by the encoder B, computing its $\mu$ and $\sigma$ values, and using them to normalize the embedding of A. However, this is not a good idea, because depending on which patch is sampled from B, we may end up generating fake A having a different style in each run. We have exactly the same problem when generating fake B. We want SemI2I to generate exactly the same fake data every time when we test it. To do this, we estimate the global $\mu$ and $\sigma$ values for the embeddings of both domains via Alg.~\ref{alg:moving_averages} during the training. In Alg.~\ref{alg:moving_averages}, d\_rate is a parameter ranging between 0 and 1. Note that this parameter needs to be set to a value that is close to 1 (e.g., 0.95), so that the current patches would not change the global mean and standard deviation values too much. When generating fake A, we use $\mu_{gB}$ and $\sigma_{gB}$. Similarly, we use $\mu_{gA}$ and $\sigma_{gA}$ while generating fake B.
\begin{algorithm}
\SetAlgoLined
\SetKwInOut{Input}{input}\SetKwInOut{Output}{output}
\Output{global mean $\mu_{g*}$ and std. $\sigma_{g*}$ for the embed.}
$\mu_{g*} \leftarrow$ 0, $\sigma_{g*} \leftarrow$ 0

$\text{d\_rate} \leftarrow$ 0.95 \tcp*{decay rate}

\ForEach{training iteration}{
~~sample a patch, compute $\mu_*$ and $\sigma_*$ of its emb.
$\mu_{g*} \leftarrow \text{d\_rate} \times \mu_{g*} +  \left( 1 - \text{d\_rate} \right) \times \mu_{*}$ 
$\sigma_{g*} \leftarrow \text{d\_rate} \times \sigma_{g*} +  \left( 1 - \text{d\_rate} \right) \times \sigma_{*}$ }

\caption{Estimation of the global $\mu$ and $\sigma$ values for the embeddings of both domains. $*$ corresponds to a domain (either A or B).}\label{alg:moving_averages}
\end{algorithm}
%
%
%
%

\vspace{-7mm}
\section{Experiments and Conclusion}
\vspace{-2mm}

We use Pl{\'e}iades images collected from \textit{Bad Ischl} and \textit{Villach} in \textit{Austria}, covering $27.71$ km\textsuperscript{2} and $43.59$ km\textsuperscript{2}, respectively. The images contain RGB color channels, and their resolution is 1 m. The annotations for \textit{building}, \textit{road}, and \textit{tree} classes have been manually prepared. We perform city-to-city domain adaptation. In the first experiment, we use \textit{Bad Ischl} as training and \textit{Villach} as test data. In the second experiment, we switch the training and the test data. 

\begin{figure}
\centering
\subfigure[\textit{Bad Ischl}]{%
    \includegraphics[width=0.49\linewidth]{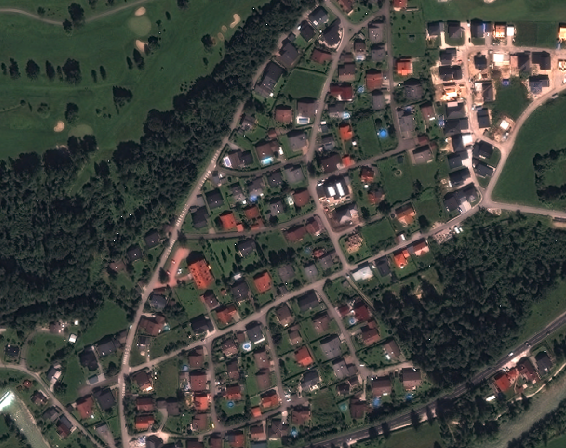}}
\subfigure[\textit{Villach}]{%
    \includegraphics[width=0.49\linewidth]{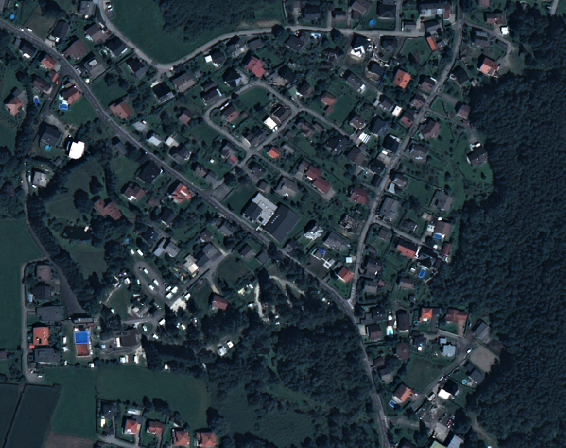}}
\subfigure[\textit{Villach} stylized fake \textit{Bad Ischl} generated by SemI2I]{%
    \includegraphics[width=0.49\linewidth]{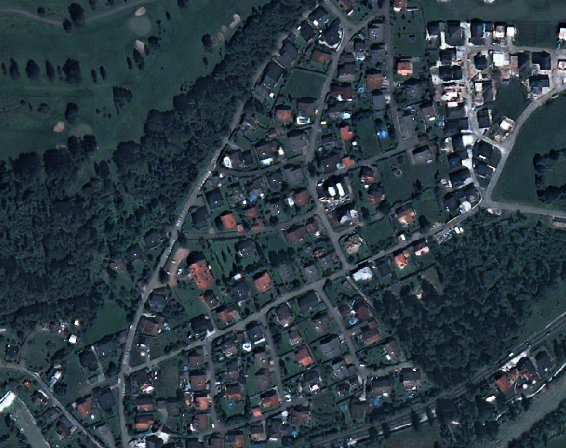}}
\subfigure[\textit{Bad Ischl} stylized fake \textit{Villach} generated by SemI2I]{%
    \includegraphics[width=0.49\linewidth]{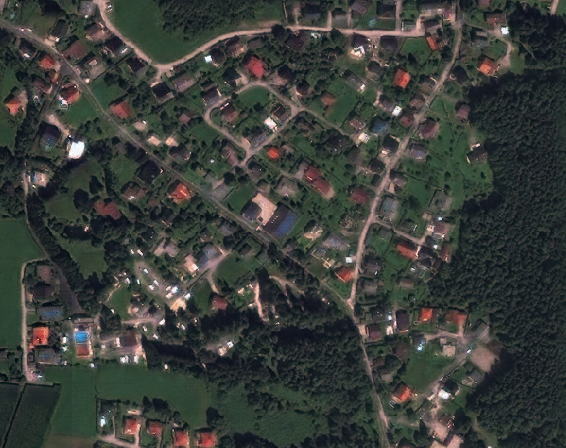}}
\subfigure[Ground-truth for (a)]{%
    \includegraphics[width=0.49\linewidth]{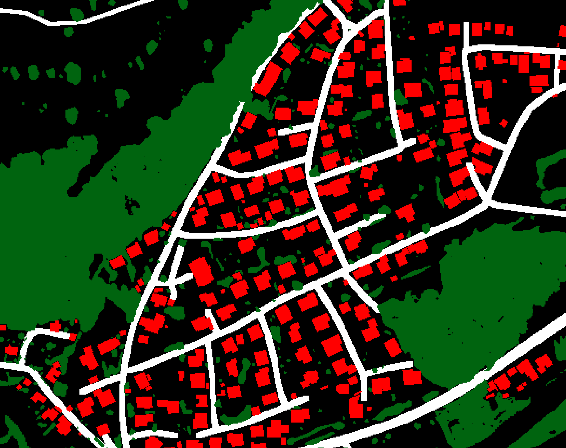}}
\subfigure[Ground-truth for (b)]{%
    \includegraphics[width=0.49\linewidth]{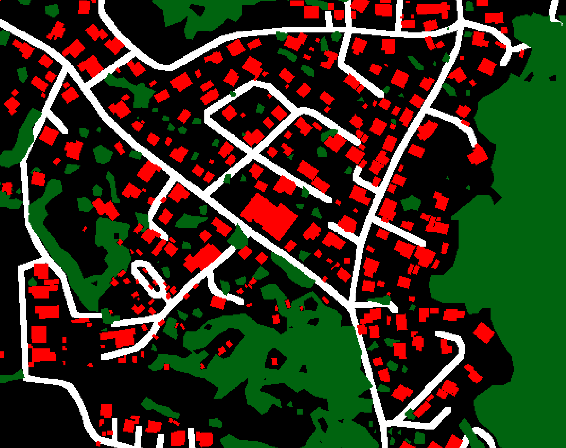}}
\subfigure[Preds. for (a) by U-net]{%
    \includegraphics[width=0.49\linewidth]{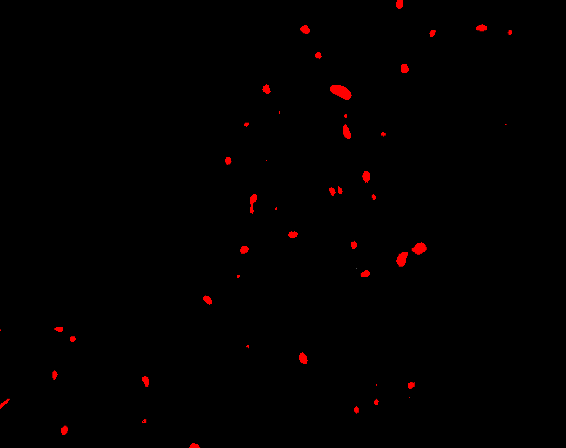}}
\subfigure[Preds. for (a) by our framework]{%
    \includegraphics[width=0.49\linewidth]{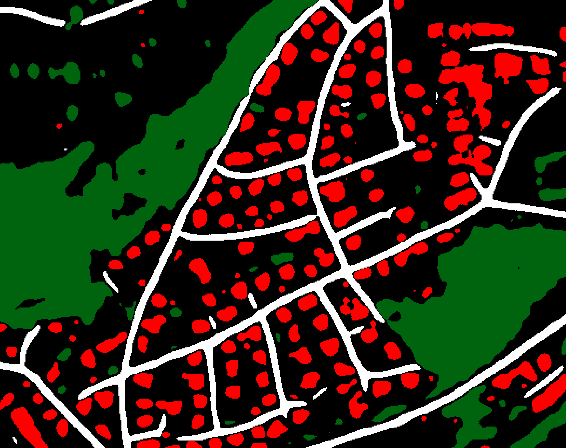}}
\subfigure[Preds. for (b) by U-net]{%
    \includegraphics[width=0.49\linewidth]{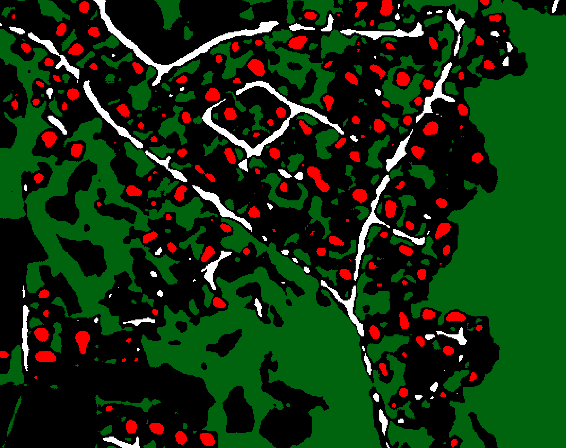}}
\subfigure[Preds. for (b) by our framework]{%
    \includegraphics[width=0.49\linewidth]{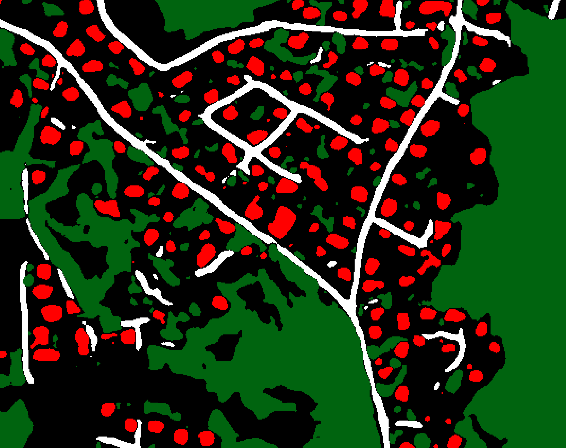}}
\caption{Training and test data used in both experiments, their ground-truth, test stylized fake training images generated by SemI2I, and the predictions generated by U-net and our framework. \textit{Red, white, } and \textit{green} colors represent \textit{building, road}, and \textit{tree} classes, respectively.}
\label{fig:results}
\end{figure}

\begin{table*}
\centering
\caption{IoU scores.}
\label{table:ious}
\scalebox{0.78}{
\begin{tabular}{||p{0.01cm}p{0.1cm}|c||c|c|c|c||c|c|c|c||}
\hline			
\multicolumn{3}{||c||}{\multirow{2}{*}{\textbf{Method}}} & \multicolumn{4}{c||}{\textbf{Training: Bad Ischl, Test: Villach}} & \multicolumn{4}{c||}{\textbf{Training: Villach, Test: Bad Ischl}} \\
\cline{4-11}                         
& \multicolumn{2}{c||}{} & \textbf{building} & \textbf{road} & \textbf{tree} & \textbf{Overall} 
& \textbf{building} & \textbf{road} & \textbf{tree} & \textbf{Overall} \\ 
\hline
\multicolumn{3}{||c||}{U-net}                       & 23.61 &  0.91 & 40.53 & 21.68 & 5.84 &  0.24 &  0.50 &  2.19  \\
\hline
\multicolumn{3}{||c||}{AdaptSegNet Single\cite{tsai2018learning}} &  6.01 &  4.37 & 10.43 &  6.94 &  3.06 &  2.71 & 10.23 &  5.33 \\
\multicolumn{3}{||c||}{AdaptSegNet Multi\cite{tsai2018learning}}  & 24.59 &  9.02 & 56.08 & 29.86 & 14.26 &  4.46 & 24.66 & 14.46 \\
\hline
\multirow{7}{*}{\rotatebox{90}{Our proposed}}& \multirow{7}{*}{\rotatebox{90}{framework with}}& CycleGAN\cite{zhu2017unpaired}
                                                    & 43.03 & 28.96 & \textbf{68.86} & 46.95 & 43.62 & 38.69 & 71.68 & 51.33 \\
& & UNIT\cite{liu2017unsupervised}                  & 30.86 & 15.84 & 63.00 & 36.57 & 19.29 & 36.83 & 35.57 & 30.56 \\
& & MUNIT\cite{huang2018multimodal}                 &  0.02 &  1.38 & 47.23 & 16.21 &  6.20 &  0.13 &  0.05 &  2.13 \\
& & DRIT\cite{lee2018diverse}                       &  0.01 &  3.96 &  8.72 &  4.23 &  0.00 & 10.19 &  0.01 &  3.40 \\
& & Gray world\cite{buchsbaum1980spatial}           & 25.19 & 26.43 & 56.15 & 35.92 & 29.55 & 24.80 & 46.41 & 33.58 \\
& & Histogram matching\cite{Gonzalez}               & 24.95 & 29.34 & 61.59 & 38.63 &  6.45 &  0.92 &  1.28 &  2.88 \\
& & ColorMapGAN\cite{tasar2019colormapgan}          & \textbf{48.47} & 37.82 & 58.92 & 48.40 & 49.16 & 41.75 & 59.84 & 50.25 \\
& & SemI2I (ours)                                   & 47.79 & \textbf{38.22} & 68.76 & \textbf{51.59} & \textbf{53.38} & \textbf{47.59} & \textbf{79.17} & \textbf{60.05} \\
\hline  
\end{tabular}}
\end{table*}

We split the cities into 256$\times$256 patches with an overlap of 32 pixels. In each training iteration, we randomly sample only 1 patch from each domain. We train SemI2I for 25 epochs, where the number of iterations in each epoch is the minimum of the number of patches from each domain. We optimize SemI2I with Adam optimizer. We set the learning rate to 0.001 in the first 15 epochs, and compute it in the rest of the epochs as:
\begin{equation}\label{eq:lr}
\text{LR} = 0.001 \times \frac{\text{num\_epochs} - \text{epoch\_no}}{\text{num\_epochs} - \text{decay\_epoch}}, 
\end{equation}
where LR, num\_epochs, epoch\_no are the current learning rate, the total number of epochs, and current epoch no. decay\_epoch stands for the epoch, which is set to 15 in our experiments, where the learning rate is started to be reduced. We set $\lambda_1, \lambda_2, \lambda_3$ and $\lambda_4$ in Eqs.~\ref{eq:g_loss} and \ref{eq:d_loss} to 10, 10, 10, and 1, respectively. We have found these values empirically. To generate maps, we first train a U-net for 8,000 iterations on the real training data. We then fine-tune it using the fake image patches generated by SemI2I and the original ground-truth for 2,500 iterations. In each training iteration of both the initial training and the fine-tuning steps, we sample a mini-batch of 32 patches. We optimize U-net in both steps via Adam optimizer with the learning rate of 0.0001.


Among the I2I approaches, we compare SemI2I with CycleGAN~\cite{zhu2017unpaired}, UNIT~\cite{liu2017unsupervised}, MUNIT~\cite{huang2018multimodal}, DRIT~\cite{lee2018diverse}, histogram matching~\cite{Gonzalez}, and gray world algorithm~\cite{buchsbaum1980spatial}. To make a fair comparison, we replace the fake images used in the fine-tuning step by the fake images generated by these methods. We also provide the results for the traditional U-net~\cite{ronneberger2015u} without doing domain adaptation and AdaptSegNet~\cite{tsai2018learning}, which aims at adapting the model instead of modifying the data. 

Fig.~\ref{fig:results} depicts the close-ups from the city pair used in the experiments, test stylized fake training images generated by SemI2I, the ground-truth, and the predictions by U-net and our framework. As can be seen, our method performs the style transfer perfectly, and real and the fake images are semantically consistent. The figure shows that U-net performs extremely poorly, when there is a large shift between the data distributions of training and test images. The improvement of our framework against U-net can clearly be observed. The quantitative results for our segmentation framework and for the other methods are reported in Table~\ref{table:ious}, which demonstrates the superior performance of our methodology.

In this work, we presented our novel SemI2I approach, which is a new data augmentation method. We compared our segmentation framework with no less than 10 different methods and showed that it exhibits a much better performance than the existing domain adaptation approaches. In the future, we plan to tackle multiple cities-to-multiple cities adaptation problem instead of performing city-to-city adaptation.

\vspace{-5mm}
\bibliographystyle{IEEEbib}
\bibliography{refs}

\begin{thebibliography}{10}

\bibitem{ronneberger2015u}
O.~Ronneberger, P.~Fischer, and T.~Brox,
\newblock ``U-net: Convolutional networks for biomedical image segmentation,''
\newblock in {\em MICCAI}, 2015.

\bibitem{khalel2019multi}
A.~Khalel, O.~Tasar, G.~Charpiat, and Y.~Tarabalka,
\newblock ``Multi-task deep learning for satellite image pansharpening and
  segmentation,''
\newblock in {\em IGARSS}, 2019.

\bibitem{tasar2019incremental}
O.~Tasar, Y.~Tarabalka, and P.~Alliez,
\newblock ``Incremental learning for semantic segmentation of large-scale
  remote sensing data,''
\newblock {\em IEEE JSTARS}, 2019.

\bibitem{tasar2019continual}
O.~Tasar, Y.~Tarabalka, and P.~Alliez,
\newblock ``Continual learning for dense labeling of satellite images,''
\newblock in {\em IGARSS}, 2019.

\bibitem{Gonzalez}
R.~C. Gonzalez and R.~E. Woods,
\newblock {\em Digital Image Processing (3rd Edition)},
\newblock 2006.

\bibitem{buchsbaum1980spatial}
G.~Buchsbaum,
\newblock ``A spatial processor model for object colour perception,''
\newblock {\em Journal of the Franklin institute}, 1980.

\bibitem{zhu2017unpaired}
J.-Y. Zhu, T.~Park, P.~Isola, and A.~A. Efros,
\newblock ``Unpaired image-to-image translation using cycle-consistent
  adversarial networks,''
\newblock in {\em CVPR}, 2017.

\bibitem{liu2017unsupervised}
M.-Y. Liu, T.~Breuel, and J.~Kautz,
\newblock ``Unsupervised image-to-image translation networks,''
\newblock in {\em NIPS}, 2017.

\bibitem{huang2018multimodal}
X.~Huang, M.-Y. Liu, S.~Belongie, and J.~Kautz,
\newblock ``Multimodal unsupervised image-to-image translation,''
\newblock in {\em ECCV}, 2018.

\bibitem{lee2018diverse}
H.-Y. Lee, H.-Y. Tseng, J.-B. Huang, M.~K. Singh, and M.-H. Yang,
\newblock ``Diverse image-to-image translation via disentangled
  representations,''
\newblock in {\em ECCV}, 2018.

\bibitem{tsai2018learning}
Y.-H. Tsai, W.-C. Hung, S.~Schulter, K.~Sohn, M.-H. Yang, and M.~Chandraker,
\newblock ``Learning to adapt structured output space for semantic
  segmentation,''
\newblock in {\em CVPR}, 2018.

\bibitem{tasar2019colormapgan}
O.~Tasar, S~L Happy, Y.~Tarabalka, and P.~Alliez,
\newblock ``{ColorMapGAN}: Unsupervised domain adaptation for semantic
  segmentation using color mapping generative adversarial networks,''
\newblock {\em arXiV}, 2019.

\bibitem{huang2017arbitrary}
X.~Huang and S.~Belongie,
\newblock ``Arbitrary style transfer in real-time with adaptive instance
  normalization,''
\newblock in {\em ICCV}, 2017.

\bibitem{mao2017least}
X.~Mao, Q.~Li, H.~Xie, Raymond~Y.K. Lau, Z.~Wang, and S.~P. Smolley,
\newblock ``Least squares generative adversarial networks,''
\newblock in {\em ICCV}, 2017.

\end{thebibliography}

\end{document}